\documentclass[pra,floatfix,twocolumn,superscriptaddress,showpacs,preprintnumbers,longbibliography,nofootinbib]{revtex4-1}
\usepackage[utf8]{inputenc} 
\usepackage{bibunits} 
\usepackage{amssymb,amsfonts,amsmath}
\usepackage{float} 
\usepackage{amsthm}
\usepackage{enumitem}
\usepackage[mathscr]{euscript}%
\usepackage{bbold}
\usepackage[normalem]{ulem}
\usepackage{hyperref} 
\hypersetup{colorlinks=true, urlcolor=blue, citecolor=blue,linkcolor=blue}
\usepackage[all]{hypcap} 

\usepackage{graphicx}
\usepackage{dcolumn}
\usepackage{bm}
\usepackage{psfrag}
\usepackage{epsfig}
\usepackage{multirow}
\usepackage{color}
\usepackage{bbm}
\usepackage[FIGTOPCAP,raggedright,nooneline]{subfigure}
\usepackage{verbatim}
\usepackage{upgreek} 
\usepackage{physics}
\usepackage{tikz} 

\newcommand{\T}[1]{\text{#1}}

\newcommand{\ignore}[1]{}

\begin{document}

		\title{Non-Hermitian skin effect as an impurity problem}
	
		\author{Federico Roccati}
		\affiliation{Universit$\grave{a}$  degli Studi di Palermo, Dipartimento di Fisica e Chimica -- Emilio Segr$\grave{e}$, via Archirafi 36, I-90123 Palermo, Italy}
		
		\begin{abstract}
			A striking feature of non-Hermitian  
			tight-binding Hamiltonians is the high sensitivity of both spectrum and eigenstates to  boundary conditions. Indeed, if the spectrum under periodic boundary conditions is point gapped, by opening the lattice the non-Hermitian skin effect 
			will necessarily occur.  
			Finding the exact skin eigenstates may be demanding in general, and many methods in the literature are based on ansatzes and on recurrence equations for the eigenstates' components. Here we devise a general procedure based on the Green's function method to calculate the eigenstates of  non-Hermitian tight-binding Hamiltonians under open boundary conditions. We apply it to the Hatano-Nelson and non-Hermitian SSH models and 
			finally we contrast the edge states localization with that of bulk states.
		\end{abstract}
		
		\maketitle

	\section{Introduction}
	For more than twenty years, since the seminal paper~\cite{benderPRL1998} in 1998, there has been a growing interest in the field of non-Hermitian quantum mechanics~\cite{ashida2020nonhermitian,benderRPP2007,bagarello2015non} and whether the full Hamiltonian of a quantum system has to be Hermitian is source of debate~\cite{mostafazadehIJGMMP2010,benderAJoP2003}. 
	From the physical standpoint, 
	non-Hermitian Hamiltonians are 
	often understood as a tool to make an effective description of the evolution of a quantum system that interacts with an environment, where quantum jumps are discarded~\cite{ArkhipovPRA2020,mingantiPRA2020}. Therefore these \textit{effective} Hamiltonians, do not generate the full quantum dynamics, but provide a correct description of an open dynamics as long as stochastic jumps can be avoided~\cite{chen2021quantum,naghiloo2019quantum}.
	
	Peculiar feature of 
	non-Hermitian Hamiltonians is the existence of Exceptional Points (EPs), that are points in parameter space at which spectrum \textit{and} eigenstates become degenerate and therefore diagonalizability is lost~\cite{heissJPAMT2012,Mirieaar7709}. One of the main quest is how these EPs can be harnessed, and many efforts have already been made~\cite{WiersigPR2020,BudichPRL2020,garmon2021anomalousorder,mcdonald2020exponentially,roccatiQST2021,quirozPR2019,lauNC2018}.
	
	One of the most powerful tools in the study of periodic systems (solids, coupled cavity arrays,$\ldots$) has certainly been the understanding of  band structure topology and its connection to the existence of topologically robust edge states~\cite{HasanRMP2010,AltlandPRB1997}. This has culminated in the celebrated bulk-boundary correspondence (BBC): the presence of boundary modes can be predicted by a topological number that depends only on bulk modes~\cite{HasanRMP2010}. This correspondence is based on the tacit assumption that, as long as the system is large, boundary conditions do not affect bulk properties. 

	Natural questions which have been investigated are related to
	whether non-Hermiticity disrupts topological properties~\cite{BergholtzRMP2021,alvarez2018topological}, whether new topological invariants can be introduced~\cite{GongPRX2018,KawabataPRX2019,wanjura2020topological}, and whether BBC holds true and in which sense~\cite{KunstPRL2018,EdvardssonPRR2020,LeePRL2016,xiong2018does}.
	 
\begin{figure}
		\centering
		\includegraphics[width=\linewidth]{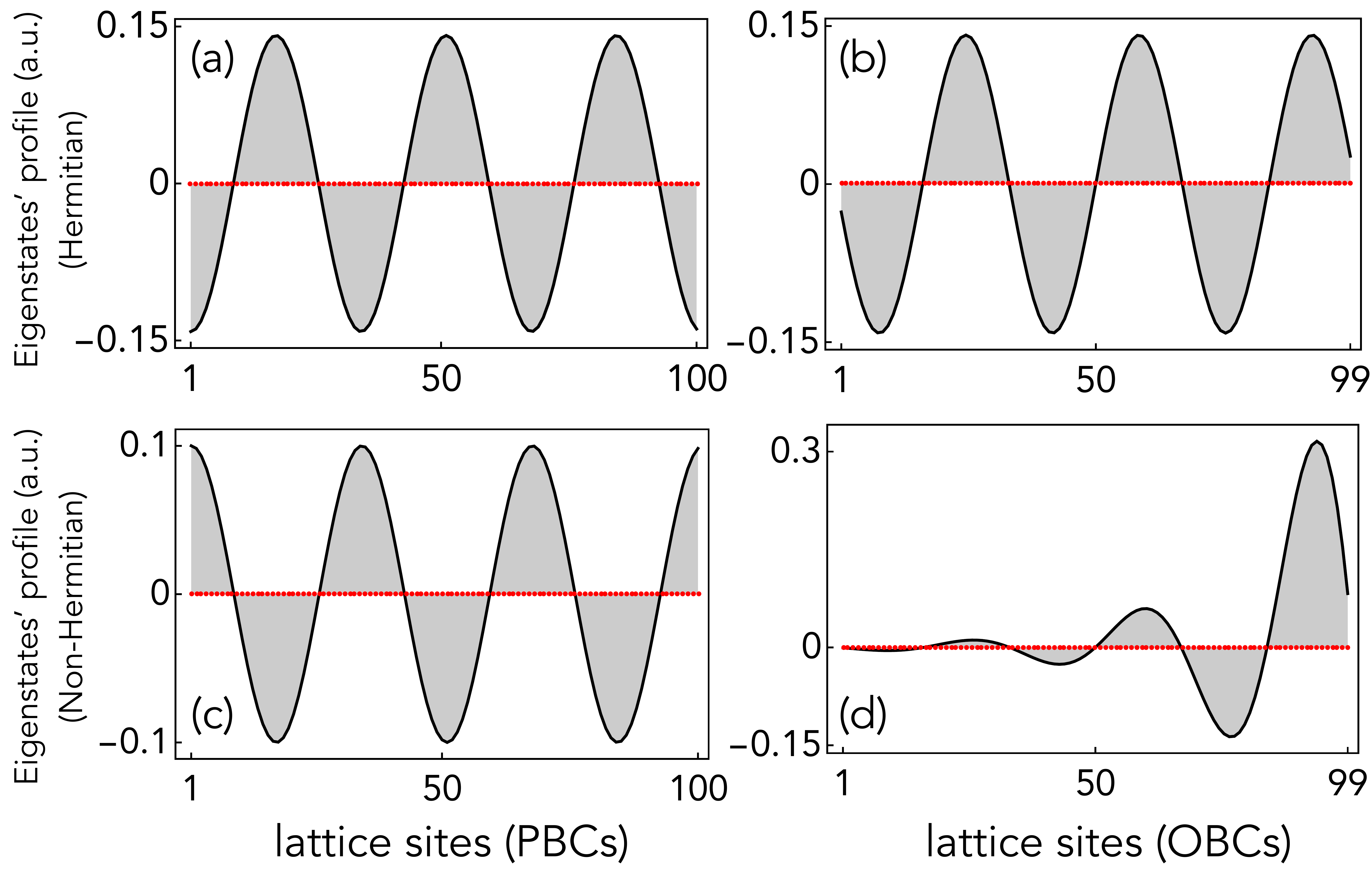}
		\caption{
			Non-Hermitian skin effect in the simplest lattice with point gapped PBC spectrum, the Hatano-Nelson model, fig.~\ref{setups}(d), eq.~\eqref{HatanoPBCS}. In all panels we plot one representative eigenstate (black), as all of them have the same qualitative behavior. In red are highlighted the lattice sites, 100 under PBCs and one less under OBCs as the chain is opened placing an infinite potential in the last site (that is decoupling it from the others). Under PBCs (left column), both in the Hermitian (a, $\delta=0$) and in the non-Hermitian case (b, $\delta=1/20$) eigenstates are delocalized. Under OBCs (right column), in the Hermitian case (c, $\delta=0$)  eigenstates are delocalized while in the non-Hermitian case (d, $\delta=1/20$) they all accumulate at the right boundary.
		}\label{skin_effect}
	\end{figure}

	A major issue regarding the restoration a non-Hermitian BBC is that 
	non-Hermitian 1D tight-binding Hamiltonians with point gapped spectrum under periodic boundary conditions (PBCs) always yield the non-Hermitian skin effect~\cite{OkumaPRL2020,YaoPRL2018,SongPRL2019,li2020critical,OkugawaPRB2020,KawabataPRB2020,LonghiPRB2020,zhang2021universal,mcdonald2021nonequilibrium,LonghiPRB2021}, that is the unusual accumulation of bulk eigenstates at the ends of the same lattice under open boundary conditions (OBCs), fig.~\ref{skin_effect}.
	This phenomenon has no Hermitian counterpart, as under OBCs the eigenstates of an Hermitian lattice Hamiltonian are delocalized wave functions. Along with the skin effect, high spectral sensitivity with respect to boundary conditions is typical of non-Hermitian Hamiltonians with point gap spectrum~\cite{KawabataPRX2019}, that is when the latter is represented by a closed curve in complex plane. This is precisely why, when dealing with the spectrum of a non-Hermitian lattice Hamiltonian, one needs to specify the boundary conditions. In particular, many models of interest display an entirely real spectrum under OBCs and a complex one under PBCs. This transition is often understood, as in $\mathcal{PT}$-symmetric models, in terms of the crossing of one or more high order exceptional points in the moving from PBCs to OBCs~\cite{BergholtzRMP2021}. This can be achieved by tuning one or more couplings in the lattice such that setting their values to zero one effectively opens the chain, leaving unchanged the number of sites. This approach has been widely investigated in the literature~\cite{guo2021exact,KunstPRL2018,YaoPRL2018} and allows the study of the spectrum's and eigenstates' transition from PBCs to OBCs. In particular, in this approach, the calculation of the eigenstates under open or generalized boundary conditions is based on an ansatz on their profile. This leads to recurrence equations for the components of their wave functions~\cite{guo2021exact}.

	In this work, we propose a complementary approach based on the Green's function method that provides a model-independent procedure to diagonalize a non-Hermitian lattice Hamiltonian, highlighting a more physical origin of skin states. 
	We will describe the method in full generality and then apply it to the relevant cases on the Hatano-Nelson and the non-Hermitian SSH model.
	
	\section{From periodic to open boundary conditions}
	Consider a general one dimensional non-Hermitian bosonic lattice Hamiltonian with point gap spectrum~\cite{BergholtzRMP2021} under PBCs with $N$ unit cells and $N_s$ sublattices where we only allow
	for hoppings within the same unit cell or between nearest-
	neighbor unit cells
	\begin{equation}
	H=\sum_{m,n=1}^N \sum_{\alpha,\beta=1}^{N_s} J_{mn}^{\alpha\beta}\dyad{m,\alpha}{n,\beta} 
	\end{equation}
	where 
	$\ket{N+1,\alpha}\equiv\ket{1,\alpha}$ for all necessary $\alpha$'s to ensure periodic boundary conditions. Such tight binding Hamiltonians occur in very different fields~\cite{LonghiPRA2016,CiccarelloPRA2011,bagarello2019tridiagonality,SanchezPRA2020}  and can be implemented in several platforms~\cite{OzawaRMP2019}. We assume uniform on-site energies, which are therefore set to zero. This Hamiltonian can be, at least in principle, exactly diagonalized through the Bloch's theorem as the system is translationally invariant under PBCs~\cite{SanchezPRA2020}. If the number of sublattices $N_s$ is less or equal to 3, which in most relevant models is always the case, then the eigenvalues $E_{\alpha}(k)$, where $k\equiv k_q=2\pi q/N$, $q=1,\ldots,N$ and the band index $\alpha=1,\ldots,N_s$, can be worked out exactly and bulk eigenstates are given by superposing the sublattices Bloch sums. 
	
\begin{figure}
		\centering
		\includegraphics[width=\linewidth]{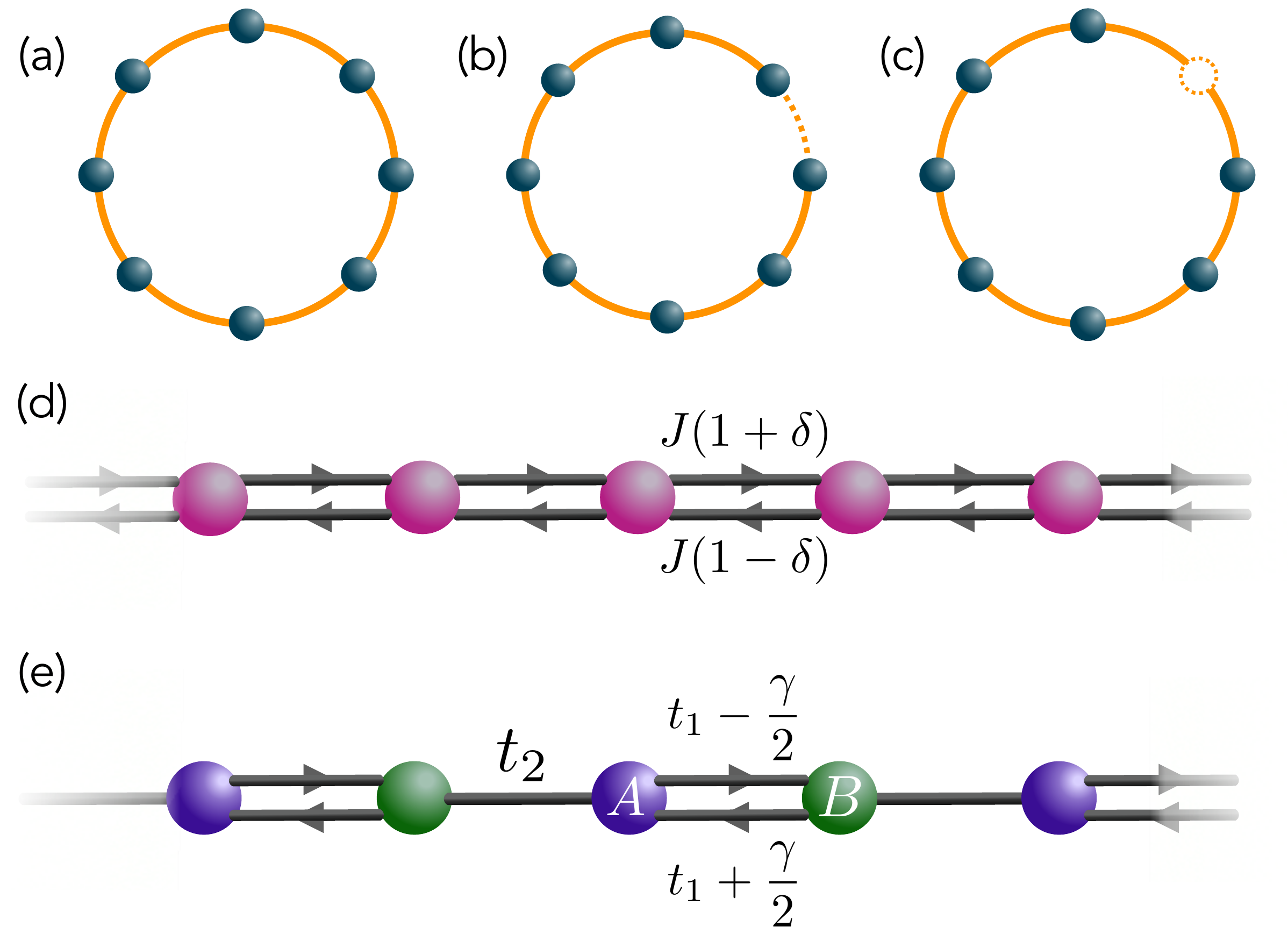}
		\caption{
			(a): General one dimensional (simple) lattice under periodic boundary conditions. These can be opened at least in two ways: by tuning one of the couplings (b), or, as proposed in this work, placing an infinite potential (i.e.~a vacancy) on one site (c) reducing by one the total number of sites.
			(d): 
			Hatano-Nelson model. 
			(e): Non-Hermitian SSH model.}\label{setups}
	\end{figure}

	In order to study the transition from PBCs to OBCs, one method~\cite{guo2021exact} is to tune the couplings $J_{1,N}^{\alpha\beta}$ and $J_{N,1}^{\alpha\beta}$ between first and last cells by replacing them with $\delta_{R}^{\alpha\beta} J_{1,N}^{\alpha\beta}$ and $ \delta_{L}^{\alpha\beta}J_{N,1}^{\alpha\beta}$ where $\delta_{R/L}^{\alpha\beta}\in[0,1]$ are tunable parameters that interpolate among PBCs (when 1) and OBCs (when 0), see fig.~\ref{setups}(a-b).
	
	There are many ways in which OBCs can be determined, as one can set all $\delta_{R/L}^{\alpha\beta}=0$ for all $\alpha,\beta$'s, or, if the hoppings inside a unit cell are only nearest neighbors, by setting $\delta_{R/L}^{\alpha\beta}=0$ for \textit{some} $\alpha,\beta$'s. The former method  effectively decouples one cell from the lattice, yielding a lattice with $N-1$ cells under OBC, while the latter yields a lattice under OBCs with $N-1$ cells \textit{plus} a broken cell. Under these generalized boundary conditions it is possible, at least for some models, to study the spectrum and profile of the eigenstates. Usually an ansatz is made for the wave function and recurrence equations for its components need to be solved~\cite{guo2021exact,KunstPRL2018,YaoPRL2018}.  
	
	 In this work we propose a different approach  to study the PBCs$-$OBCs transition, which allows in full generality the calculation of OBC spectrum and skin states without any specific ansatz. Instead of considering tunable \textit{couplings}, in order to open the chain, we consider tunable \textit{local impurities}, fig.~\ref{setups}(a,c). The Hamiltonian under PBCs, but no anymore translationally invariant, with local impurities reads
	\begin{equation}
	H^{\varepsilon_\alpha} = H+\sum_\alpha \varepsilon_\alpha \dyad{N,\alpha}
	\end{equation}
	where $\varepsilon_\alpha>0$. This Hamiltonian describes the same system with a set of potential barriers on (generally all) the sites of last cell. In the limit $\varepsilon_\alpha\rightarrow\infty$ for all $\alpha$'s, all corresponding lattice sites decouple, effectively opening the chain: 
	\begin{equation}
	\tilde H = 
	\lim_{ \varepsilon_\alpha\rightarrow\infty}H^{\varepsilon_\alpha}
	\end{equation}
	where $\tilde H$ denotes the Hamiltonian under OBCs. The main advantage of opening the chain through local impurities instead of tunable couplings, is that 
	the Green's function of a lattice Halitonian with impurities 
	can be exactly calculated~\cite{economou2006green}, even for finite $\varepsilon_\alpha$, and this allows the calculation of the eigenvalues (poles of the Green's function) and of the corresponding impurity states. We will show that, for two representative models, these impurity states are either generalization of edge/vacancy states (if any), and skin states coming from the bulk. The localization of bulk skin states will result  from our method without any ansatz on their profile.
	
	In a Hermitian lattice in thermodynamic limit $N\rightarrow\infty$, the presence of one finite impurity adds one pole to the Green's function leaving the continuous spectrum unmodified and the eigenstate corresponding to this additional pole is localized around the impurity~\cite{economou2006green}. For a finite but large Hermitian lattice instead, the part of the spectrum that becomes continuous in the thermodynamic limit, along with the corresponding unbound modes, is slightly perturbed, except for the new pole still corresponding to a bound state near the impurity. Furthermore, in the limit where OBCs are achieved ($\varepsilon_\alpha$ $\rightarrow\infty$), one is left with a shorter Hermitian lattice with slightly perturbed spectrum and delocalized normal modes.
	
	Remarkably, for the non-Hermitian lattice  of our interest, under OBCs, the poles of the Green's function are \textit{all} drastically different from those of the  lattice under PBCs. All the corresponding impurity bulk states are the so called skin states which are too drastically different from the delocalized normal modes of the lattice under PBCs, showing that the skin effect can be understood in terms of an impurity problem.
	More concretely, considering only one impurity in the last site of a simple lattice ($N_s=1$)  $\ket{N}$  the corresponding Green's  function defined by $G^{\varepsilon}(z)=(z-H^{\varepsilon})^{-1}$ reads~\cite{economou2006green,LeonfortePRL2021,LombardoPRA2014}
	\begin{equation}\label{GreenNeps}
	G^{\varepsilon}(z) = G(z) + \varepsilon\frac{ G(z)\dyad{N}G(z)}{1-\varepsilon\bra{N}G(z)\ket{N}}  
	\end{equation}
	where $G(z)=(z-H)^{-1}$ is the Green's function of the lattice under PBCs. In order to deal with more impurities, eq.~\eqref{GreenNeps} can be used iteratively. The poles of $G^{\infty}(z)$ are the $N-1$ eigenvalues $\tilde E$'s and under OBCs ($\varepsilon\rightarrow\infty$) the right and left (unnormalized) eigenstates are given by
	\begin{eqnarray}\label{skinR}
	\ket{\tilde \Psi^{R}(\tilde E)}
	&=&
	\sum_{n} 
	G(n,N;\tilde E)
	\ket{n}\\
	\bra{\tilde \Psi^{L}(\tilde E)}
	&=&
	\sum_{n}
	G(N,n;\tilde E)
	\bra{n}\label{skinL}
	\end{eqnarray}
	where $G(m,n;z)=\bra{m}G(z)\ket{n}$ and $n$ runs up to $N-1$, as the OBC lattice has one site less than the PBC one.

	On the one hand, if $\ket{\psi_k}$'s are the unnormalized eigenstates of a Hermitian operator $\hat H$, in order to construct the completeness relation the eigenstates are normalized as $\ket{\psi_k}\rightarrow\ket{\psi_k}/\sqrt{\braket{\psi_k}{\psi_k}}$ and this normalization coefficient is unique up to a state dependent \textit{phase} factor. On the other hand, for a non-Hermitian Hamiltonian $\hat H(\lambda)$, where $\lambda$ parametrizes non-Hermiticity, both left and right eigenstates $\ket{\psi^{R/L}_k(\lambda)}$ are needed 
	in order to form the closure relation. A possible way of having a consistent Hermitian limit as $\lambda\rightarrow0$, is to  \textit{binormalize}~\cite{brody2013biorthogonal} left and right eigenstates  as 
	\begin{eqnarray}
	\ket{\psi^{R}_k(\lambda)}
	&\rightarrow&\ket{\psi^{R}_k(\lambda)}/\sqrt{\braket{\psi^{L}_k(\lambda)}{\psi^{R}_k(\lambda)}}\\
	\bra{\psi^{L}_k(\lambda)}
	&\rightarrow&\bra{\psi^{L}_k(\lambda)}/\sqrt{\braket{\psi^{L}_k(\lambda)}{\psi^{R}_k(\lambda)}},
	\end{eqnarray}
	so that $\sum_k \dyad{\psi^{R}_k(\lambda)}{\psi^{L}_k(\lambda)}=\mathbb{1}$. However, this binormalization is unique up to a state dependent \textit{scale} factor, that is, binormalization and completeness relation are invariant under the transformations $\ket{\psi^{R}_k(\lambda)}\rightarrow A_k\ket{\psi^{R}_k(\lambda)}$ and $\bra{\psi^{L}_k(\lambda)}\rightarrow A_k^{-1}\bra{\psi^{L}_k(\lambda)}$ for any non zero $A_k$. Therefore,  we will assume in the following that right and left eigenstates are not normalized or binormalized, unless otherwise specified.

	\section{Hatano-Nelson model}
	The Hatano-Nelson model without disorder~\cite{HatanoPRL1996,BergholtzRMP2021}, fig.~\ref{setups}(d), is the prototypical example of a non-Hermitian Hamiltonian exhibiting the non-Hermitian skin effect, fig.~\ref{skin_effect}.  
	Its Hamiltonian under PBCs reads
	\begin{equation}\label{HatanoPBCS}
	H_\T{HN} =
	\sum_{n=1}^N J(1+\delta)\dyad{n+1}{n} + J(1-\delta)\dyad{n}{n+1}
	\end{equation}
	where we assume that both $J$ and $\delta$ are real, $\delta\in[-1,1]$ and  $\ket{N+1}\equiv\ket{1}$. Being a simple lattice ($N_s=1$), its  left eigenstates are obtained by Hermitian conjugation of the right ones despite the Hamiltonian being non-Hermitian
	$\ket{\Psi(k)}
	=1/\sqrt{N} 
	\sum_{n=1}^N e^{ikn}\ket{n}$
	where $k \equiv k_q= 2\pi q/N$, $ q=1,\ldots,N$ and its spectrum is given by 
	\begin{equation}
	E(k) =2J\left( \cos k -i\delta \sin k\right)
	\end{equation}
	forming an ellipse (in the thermodynamic limit) in complex plane.

\begin{figure}
		\centering
		\includegraphics[width=0.9\linewidth]{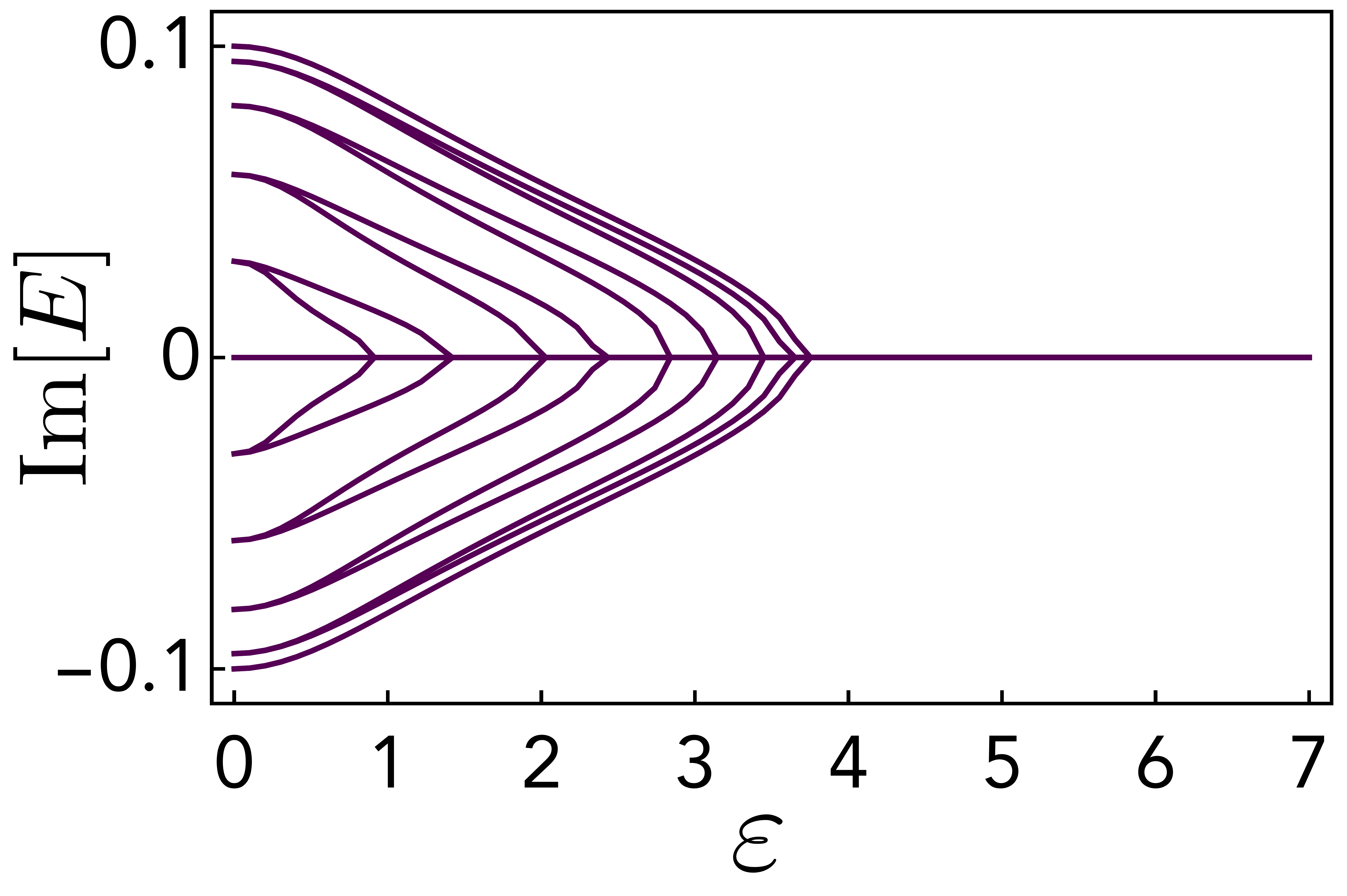}
		\caption{
			Imaginary part of the spectrum of an Hatano-Nelson lattice with 20 sites under PBCs with a local potential $\varepsilon$ on one site as function of $\varepsilon$. As the potential barrier increases many exceptional points are crossed, until the spectrum becomes purely real.}\label{passingEPs}
	\end{figure}

	We describe the transition from PBCs to OBCs as discussed in the previous section, that is tuning one local potential in the last site. Defining $H_\T{HN}^{\varepsilon} = H_\T{HN} + \varepsilon \dyad{N}{N}$, we can write the Hamiltonian under OBCs as $\tilde H_\T{HN}=H_\T{HN}^{\infty} $ (this equality is meant to hold for the upper left $(N-1)\times(N-1)$ block of both matrices). The Green's function of $H_\T{HN}^{\varepsilon} $ has exactly the same expression of eq.~\eqref{GreenNeps} and its poles are the solutions to~\cite{economou2006green}:
	\begin{equation}\label{polequation}
	\frac{1}{N}\sum_k \frac{1}{z - E(k)} = \frac{1}{\varepsilon}.
	\end{equation}
	For a general $\varepsilon$ only numerical roots of eq.~\eqref{polequation} are available, but still their dependence on $\varepsilon$ makes clear the crossing of many EPs as $\varepsilon\rightarrow\infty$, see fig.~\ref{passingEPs}.
	
	As discussed in~\cite{YaoPRL2018}, the OBC spectrum is obtained from the PBC one by applying the complex shift $k\rightarrow k/2-i\log \rho$, with $\rho=\sqrt{(1+\delta)/(1-\delta)}$:
	\begin{equation}\label{OBCspectrumHN}
	\tilde E(k/2) =2J\sqrt{1-\delta^2}\cos(k/2), 
	\end{equation}
	with $q=1,\ldots,N-1$, which by construction are solutions to eq.~\eqref{polequation}. 
	Considering that the PBCs Green's function has matrix elements~\cite{Belloeaaw0297}
		\begin{equation}
		\bra{m}G_\T{HN}(z)\ket{n} = \frac{1}{N}\sum_k \frac{e^{ik(m-n)}}{z - E(k)},
		\end{equation} 
	the corresponding unnormalized eigenstates can be obtained through the Green's function as described earlier and are explicitely given by
	\begin{eqnarray}\label{skinRR}
	\ket{S^R(k/2)}
	&=&
	\sum_{n}
	\left[\frac{1}{N}\sum_{k'} \frac{ e^{ik'n}    }{\tilde E(k/2) - E(k')}\right]
	\ket{n}
\\
	\bra{S^L(k/2)}
	&=&
	\sum_{n}
	\left[\frac{1}{N}\sum_{k'} \frac{ e^{-ik'n}    }{\tilde E(k/2) - E(k')}\right]
	\bra{n}.\label{skinLL}
	\end{eqnarray} 
	At this level the localization (or not) of these eigenstates is not clear, but, as proved in Appendix~\ref{appA}, we find that
	\begin{equation}
	\sum_{k'} \frac{ e^{\pm ik'n}    }{\tilde E(k/2) - E(k')}
	\propto
	\left[\frac{1+\delta}{1-\delta}\right]^{\pm n/2}
	\sin(\frac{kn}{2}).\label{resultHN} 
	\end{equation}
	Therefore, making no ansatz on the eigenstates' profiles, the skin effect naturally emerges by calculating the eigenstates of the Hatano-Nelson model through the Green's function method by placing an infinite potential (i.e.~a vacancy) on one site,  instead of opening the chain by tuning couplings. In particular for the Hatano-Nelson model all impurity states are skin states and eq.~\eqref{resultHN} shows that right and left skin states always accumulate at opposite ends of the open lattice.

	\section{Non-Hermitian SSH model}\label{SSH}
	
	Another instance where skin states naturally emerge through the Green's function method is the non-Hermitian SSH model~\cite{SSHPRB1980,LieuPRB2018}, fig.~\ref{setups}(e), whose Hamiltonian under PBCs is
	\begin{eqnarray}
	H_\T{SSH} 
	&=& \sum_{n=1}^{N} 
	\left(t_1+\frac{\gamma}{2}\right) \dyad{n,A}{n,B}
	+
	\left(t_1-\frac{\gamma}{2}\right) \dyad{n,B}{n,A}
	\nonumber\\
	&& +t_2
	\left(\dyad{n+1,A}{n,B} + \dyad{n,B}{n+1,A}\right)
	\end{eqnarray}
	where $N$ is the number of cells, $A,B$ label the two sublattices ($N_s=2$), $t_{1,2},\gamma$ are real
	and $\ket{N+1,A}\equiv\ket{1,A}$. 
	
	Under PBCs we can diagonalize $ H_\T{SSH} $ through the Bloch's theorem and its  Bloch Hamiltonian reads 
	\begin{equation}
	\mathcal H_\T{SSH}(k)
	=
	\begin{pmatrix}
	0  & f_{ab}(k) \\
	f_{ba}(k) & 0 \\
	\end{pmatrix}
	\end{equation}
	with $k\equiv k_q=2\pi q/N$ and $q=1,\ldots,N$, where $f_{ab}(k)=t_1+e^{-ik}t_2+\gamma/2$ and $f_{ba}(k)=t_1+e^{ik}t_2-\gamma/2$, and its right and left binormalized eigenstates are
	\begin{equation}
	\ket{v_\pm^R(k)} = \frac{1}{\sqrt{1+\frac{\omega(k)^2}{f_{ba}(k) f_{ab}(k)}}} 
	\left[
	\ket{A}
	\pm \frac{\omega(k)}{f_{ab}(k)} \ket{B}
	\right]
	\end{equation}
	and $\bra{v_\pm^L(k)}$, given by replacing kets with bras and 
	$ab$ by $ba$. 
	The spectrum under PBCs is given by $E_{\pm}(k)=\pm\omega(k)$ where $\omega(k)=\sqrt{f_{ab}(k)f_{ba}(k)}$.   Its full right and left eigenstates are	
	\begin{equation*}
	\ket{\Psi_\pm^R(k)}
	=
	\braket{A}{v_\pm^R(k)}
	\ket{\psi_A(k)}
	+
	\braket{B}{v_\pm^R(k)}
	\ket{\psi_B(k)}
	\end{equation*}
	\begin{equation*}
	\bra{\Psi_\pm^L(k)}
	=
	\braket{v_\pm^L(k)}{A}
	\bra{\psi_A(k)}
	+
	\braket{v_\pm^L(k)}{B}
	\bra{\psi_B(k)}
	\end{equation*}
	where $\ket{\psi_C(k)}=1/\sqrt{N} \sum_{n=1}^{N} e^{ikn}\ket{n,C}$, $C=A,B$. 
	Its Green's function matrix elements are given by
	\begin{equation*}
	\mathcal{G}_{mn}^{\alpha\beta}(z)
	=
	\frac{1}{N}
	\sum_k
	\sum_{s=\pm}
	\braket{\alpha}{v_s^R(k)}
	\braket{v_s^L(k)}{\beta}
	\frac{e^{i k (m-n)}}{z-E_{s}(k)}
	\end{equation*}
	where $\mathcal{G}_{mn}^{\alpha\beta}(z)
	=
	\bra{m,\alpha}G_\T{SSH}(z)\ket{n,\beta}$,  $m,n=1,\ldots,N$ and $\alpha,\beta=A,B$. 
	
	We now consider the transition to open boundary conditions by placing a potential on the last \textit{site} of the lattice. We define again $H_\T{SSH}^\varepsilon = H_\T{SSH}+\varepsilon \dyad{N,B}$ so that the Hamiltonian under OBCs is $\tilde H_{\T{SSH}}=H_{\T{SSH}}^{\infty}$.
	The OBCs eigenvalues are again given by the solutions to
	$\mathcal{G}_{NN}^{BB}(z)=0$ and are 
	$\tilde E_0=0$ and 
	\begin{eqnarray}
	\tilde E_\pm( k/2) &=& \pm \sqrt{ c^2 +t_2^2 +2 ct_2\cos (k/2) }
	\end{eqnarray}
	where now $q=1,\ldots,N-1$ and $c=\sqrt{t_1^2-\gamma^2/4}$. 
	The corresponding eigenstates given by the Green's function method, see eqs.~\eqref{skinR}-\eqref{skinL}, read
	\begin{equation}
	\ket{\tilde \Psi^{R}(\tilde E)}
	=
	\sum_{n=1}^{N}
	\mathcal{G}_{nN}^{AB}(\tilde E)\ket{n,A}
	+
	\sum_{n=1}^{N-1}
	\mathcal{G}_{nN}^{BB}(\tilde E)\ket{n,B}
	\label{EigStatRSSH}
	\end{equation}
	\begin{equation}\label{EigStatLSSH}
	\bra{\tilde \Psi^{L}(\tilde E)}
	=
	\sum_{n=1}^{N}
	\mathcal{G}_{Nn}^{BA}(\tilde E)\bra{n,A}
	+
	\sum_{n=1}^{N-1}
	\mathcal{G}_{Nn}^{BB}(\tilde E)\bra{n,B}
	\end{equation}
	where $\tilde E$ is any of the OBCs eigenvalues.
	We now discuss separately the bulk eigenstates and the zero energy ones as their localization properties have different physical origins.
	
	\subsection{Bulk eigenstates}
	
	Consider $\tilde E\neq0$ belonging to the OBC spectrum of the SSH lattice and is corresponding eigenstates, given by eqs~\eqref{EigStatRSSH}-\eqref{EigStatLSSH}
	. Then, as proved in  Appendix~\ref{appC}, their components satisfy the following relations:  
	\begin{equation}\label{bulkSSHA}
	\mathcal{G}_{nN}^{AB}(\tilde E)
	=
	\frac{1}{2N}
	\sum_{k}
	\sum_{s=\pm}
	\frac{f_{ab}(k)}{E_s(k)}
	\frac{e^{ikn}}{\tilde E - E_s(k)}
	\propto
	\alpha_n(\tilde{E}) 
	r^{n-1}
	\end{equation}
	\begin{equation}\label{bulkSSHB}
	\mathcal{G}_{nN}^{BB}(\tilde E)
	=
	\frac{1}{2N}
	\sum_{k}
	\sum_{s=\pm}
	\frac{e^{ikn}}{\tilde E - E_s(k)}
	\propto
	\beta_n(\tilde{E}) 
	\tilde{E}
	r^{n}
	\end{equation}
	where $r=\sqrt{(t_1-\gamma/2)(t_1+\gamma/2)}$ and similar expressions for $\mathcal{G}_{Nn}^{BA/BB}(\tilde E)$, see Appendix~\ref{appC}. Therefore, we again see how the non-Hermitian skin effect comes up naturally from the (bulk) eigenstates calculated through the Green's function. We observe that, as in the Hatano-Nelson model, right and left bulk eigenstates always accumulate at opposite ends of the OBC lattice. Note that the components of these eigenstates on the B sublattice are proportional to the eigenvalue itself. Finally, the bulk states are localized skin states only in the non-Hermitian case ($\gamma\neq0$).
	
	\subsection{Edge states}
	
\begin{figure}
		\centering
		\includegraphics[width=\linewidth]{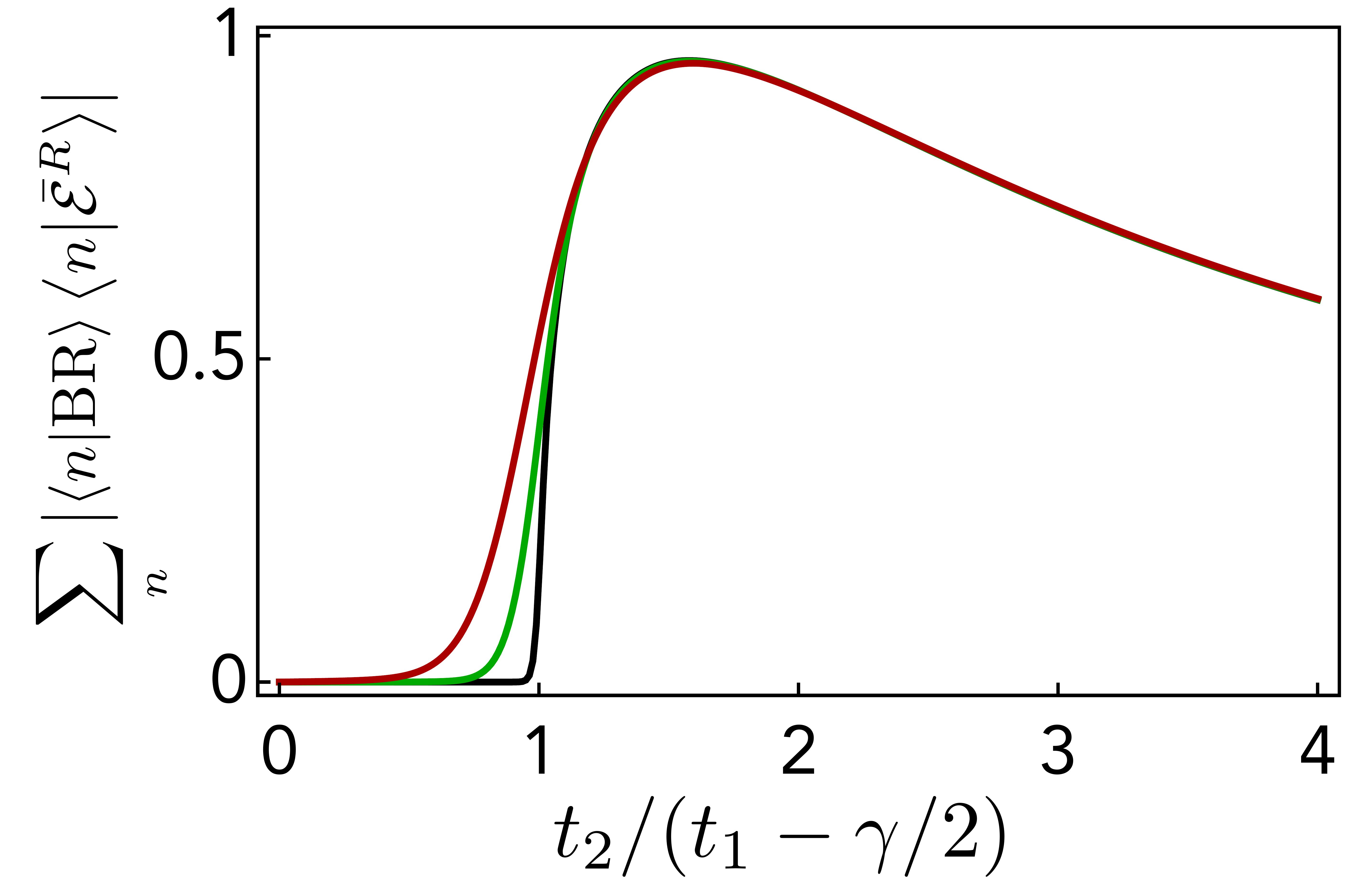}
		\caption{
			Vicinity of edge and skin states. If $t_2>t_1-\gamma/2$ ($t_2<t_1-\gamma/2$) the former localizes at the same (opposite) edge. We set $\gamma=t_1$ in the plot, as different values change only qualitatively the result. The sum is up to the number of lattice sites, that are 20 (red), 40 (green), 200 (black).}\label{proprSSH}
	\end{figure}

	The zero modes of the non-Hermitian SSH model, given   by eqs~\eqref{EigStatRSSH}-\eqref{EigStatLSSH} with $\tilde E=0$ which we label here $\ket{\mathcal E^{R/L}}$, are a generalization of the edge states of the Hermitian one~\cite{asboth2016short}.
	As the lattice is opened through one vacancy, leaving a broken cell, there will always be one topological (right) edge state localized at one of the two boundaries of the lattice.
	In particular, the  components of $\ket{\mathcal E^{R}}$ and $\bra{\mathcal E^{L}}$ are given respectively by
	\begin{equation}\label{edgeSSHresult}
	\mathcal{G}_{nN}^{AB}
	\propto
	\left[-\frac{t_1-\gamma/2}{t_2}\right]^{n},
	\quad
	\mathcal{G}_{Nn}^{BA}
	\propto
	\left[-\frac{t_1+\gamma/2}{t_2}\right]^{n}
	\end{equation}
	while $\mathcal{G}_{nN/Nn}^{BB}\equiv0$ as the $B$ components are proportional to the energy. Remarkably, these eigenstates do not follow automatically the same localization of the bulk skin states, displaying their topological robustness against the non-Hermitian skin effect. Indeed, right and left edge state can be localized on opposite or equal ends of the lattice and  non-Hermiticity can enhance or reduce their localization length. In order to find whether edge states accumulate where the bulk skin states, do we consider the quantity defined by
	\begin{equation}
	\sum_n \left| \braket{n}{\T{BR}} \braket{n}{\bar{\mathcal{E}}^R}\right|\,,
	\end{equation}
	where $\ket{\bar{\mathcal{E}}^R}$ is the \textit{normalized} right edge state and $\ket{\T{BR}}$ is the \textit{normalized} sum of all bulk right eigenstates. This number is a measure of the vicinity of these states: they localize at the same (opposite) edge if $t_2>t_1-\gamma/2$  ($t_2<t_1-\gamma/2$)  opposite edge, fig.~\ref{proprSSH}.

	\section{Conclusions}
	
	We presented a new method to calculate the eigenstates of non-Hermitian lattice Hamiltonian under open boundary conditions based on the Green's function method. Interpreting an open lattice as resulting from a closed one 
	with an added pointlike potential barrier,
	the Green's function, and therefore all eigenstates, can be exactly calculated without making any ansatz. We showed how with this method the localization of the non-Hermitian skin states emerges from the non-Hermiticity of the model, and how possible	 edge states are modified by non-Hermiticity.
	
	A tight connection between skin effect and the presence of an impurity in a one dimensional lattice through Green's function method has been discussed and recently similar tools have also been considered in relation to the skin effect~\cite{mao2021boundary,liCP2021}.
	Indeed under PBCs bulk states describe chiral propagation along the lattice
	(e.g.~$\delta$-dependent in the Hatano-Nelson model). However,
	by introducing a vacancy they are forced to accumulate close to it 
	This picture can change when considering 
	higher-dimensional systems exhibiting the skin effect~\cite{zhang2021universal,KawabataPRB2020}. 
	The generalization of the present method to such cases is under ongoing investigation.
	
	\section{Acknowledgements}
	We acknowledge fruitful discussions with Fabio Bagarello, Francesco Gargano, Salvatore Lorenzo, Angelo Carollo and Francesco Ciccarello.
	
	\appendix

	\section{Proof of eq.~\eqref{resultHN}}\label{appA}
	
	Here we derive explicitly the skin states of the Hatano-Nelson model under OBCs with $N-1$ sites whose eigenvalues are 
	\begin{equation}
	\tilde E(k/2) =2J\sqrt{1-\delta^2}\cos(k/2), 
	\end{equation}
	with $k\equiv k_q=2\pi q/N$ and  $q=1,\ldots,N-1$.  Being the eigenvalues all real and distinct, the Hamiltonian is diagonalizable and therefore all eigenspaces are non degenerate. Using the fact that the matrix representation is Toepliz, its unnormalized $k$-th (skin) right eigenstate  $\ket{\psi^R(k/2)}$ has components
	\begin{equation}
	\braket{n}{\psi^R(k/2)}=\rho^n \sin(k n/2),\quad n=1,\ldots,N-1.
	\end{equation}  
	On the other hand, through the Green's function method we know that the $k$-th (skin) right eigenstate  $\ket{\phi^R(k/2)}$ has components
	\begin{equation}
	\braket{n}{\phi^R(k/2)} = G(n,N;\tilde E(k/2)).
	\end{equation} 
	Therefore $\ket{\phi^R(k/2)}$ and $\ket{\psi^R(k/2)}$ must be proportional:
	\begin{equation}
	\frac{1}{N}\sum_{k'}^N \frac{   e^{ik'n}    }{\tilde E(k/2) - E(k')}
	= A(\rho,k_q)  
	\rho^n \sin(kn/2) 
	\end{equation}
	where $A(\rho,k_q)  $ is a proportionality constant and $\rho=\sqrt{(1+\delta)/(1-\delta)}$. Therefore, the right hand side of this equation is the inverse discrete Fourier transform of $[\tilde E(k/2) - E(k')]^{-1}$, so that transforming back we get 
	\begin{equation*}
	\frac{1}{\tilde E(k_q/2) - E(k')} = A( \rho,k_q) \sum_{{n}=1}^N  e^{-ik'n} \rho^n \sin(k_qn/2) 
	\end{equation*}
	which leads to
	\begin{equation}
	A_q( \rho,k_q) 
	=  
	\frac{1+ \rho^2}{ 2 \rho J(-1+(-1)^q \rho^{N})\sin(k_q/2)}.
	\end{equation}
	A byproduct of our proof is the summation rule  
	\begin{equation*}
	\frac{1}{N}\sum_{k'} \frac{e^{ik'n}}{\tilde E(k_q/2) - E(k')}
	= \frac{(1+ \rho^2) \rho^n \sin(k_qn/2)}{ 2 \rho J(-1+(-1)^q \rho^{N})\sin(k_q/2)} 
	\end{equation*}
	A similar argument holds for the left eigenstates, which leads to
	\begin{equation*}
	\frac{1}{N}\sum_{k'}^N \frac{   e^{-ik'n}    }{\tilde E(k_q/2) - E(k')}
	= A(1/ \rho,k_q)  
	 \rho^{-n} \sin(k_qn/2) 
	\end{equation*}

	\section{Eigenstates of the open SSH lattice through Green's function method}\label{appC}
	
	Consider an open SSH lattice as described in section~\ref{SSH} whose Hamiltonian is $\tilde H_\T{SSH}$. Through the non-unitary matrix 
	\begin{equation}
	W=\T{diag}\{1,r,r,r^2,r^2,\ldots,r^{N-1},r^{N-1}\},
	\end{equation}
	with $r=\sqrt{(t_1-\gamma/2)(t_1+\gamma/2)}$  we can find the Hamiltonian $\tilde H_\T{SSH}'= W^{-1}\tilde H_\T{SSH}W$  of an \textit{Hermitian} SSH model, if $t_1-\gamma/2>0$ (which anyway does not affect our results) with intracell and intercell hoppings $c=\sqrt{t_1^2-\gamma^2/4}$ and $d=t_2$, respectively. $\tilde H_\T{SSH}'$ and $\tilde H_\T{SSH}$ are isospectral and the right (left) eigenstates $\ket{ \psi_k^R}$ $\left(\bra{ \psi_k^L}\right)$ of $\tilde H_\T{SSH}$ are given by $\ket{ \psi_k^R}=W\ket{ \varphi_k}$ $\left(\bra{ \psi_k^L}=\bra{ \varphi_k}W^{-1}\right)$ where $\ket{ \varphi_k}$ are the eigenstates of the \textit{Hermitian} Hamiltonian $\tilde H_\T{SSH}'$. The spectrum and eigenstates of $\tilde H_\T{SSH}'$ are given by~\cite{shin_1997}
	$\tilde E_0=0$ and  
	\begin{equation}
	\tilde E_\pm( k/2) = \pm \sqrt{ c^2 +d^2 +2 cd\cos (k/2) },
	\end{equation}
	$k\equiv k_q=2\pi q/N$, $q=1,\ldots,N-1$, as in the main text and 
	\begin{equation}
	\ket{ \varphi_0}=
	\sum_{n=1}^{N}
	\left(-\frac{c}{d}\right)^n
	\ket{n,A}
	\end{equation}
	and
	\begin{equation}
	\ket{ \varphi_k}=
	\sum_{n=1}^{N}
	\alpha_{n,k}
	\ket{n,A}
	+
	\sum_{n=1}^{N-1}
	\tilde E_\pm( k/2)
	\beta_{n,k}
	\ket{n,B}
	\end{equation}
	where
	\begin{equation}
	\alpha_{n,k}
	=
	\frac{d}{c}
	\sin\left[\frac{(n-1)k}{2}\right]
	+
	\sin\left[\frac{nk}{2}\right]
	\end{equation}
	\begin{equation}
	\beta_{n,k}
	=
	\frac{1}{c}
	\sin\left[\frac{nk}{2}\right]
	\end{equation}
	therefore the right and left eigenstates of $\tilde H_\T{SSH}$ are 
	\begin{equation}
	\ket{ \psi_0^R}=
	\sum_{n=1}^{N}
	\left(-\frac{c}{d}\right)^n
	r^{n-1}
	\ket{n,A}
	\end{equation}
	\begin{equation}
	\bra{ \psi_0^L}=
	\sum_{n=1}^{N}
	\left(-\frac{c}{d}\right)^n
	r^{1-n}
	\bra{n,A}
	\end{equation}
	and 
	\begin{equation*}
	\ket{ \psi_k^R}=
	\sum_{n=1}^{N}
	\alpha_{n,k}
	r^{n-1}
	\ket{n,A}
	+
	\sum_{n=1}^{N-1}
	\tilde E_\pm( k/2)
	\beta_{n,k}
	r^{n}
	\ket{n,B}
	\end{equation*}
	\begin{equation*}
	\bra{ \psi_k^L}=
	\sum_{n=1}^{N}
	\alpha_{n,k}
	r^{1-n}
	\bra{n,A}
	+
	\sum_{n=1}^{N-1}
	\tilde E_\pm( k/2)
	\beta_{n,k}
	r^{-n}
	\bra{n,B}
	\end{equation*}
	As discussed in the main text, the eigenstates of $\tilde H_\T{SSH}$ are also given by eqs.~\eqref{EigStatRSSH}-\eqref{EigStatLSSH}. Being the spectrum non-degenerate, we have that these eigenstates calculated through different procedures need to be proportional
	\begin{equation}\label{propEdge}
	\ket{ \tilde \Psi^R(0)} = K_0(r)\ket{ \psi_0^R}
	\end{equation} 
	\begin{equation}\label{propBulk}
	\ket{ \tilde \Psi^R(\tilde E_\pm( k/2))} = K(r,k)\ket{ \psi_k^R}
	\end{equation} 
	where $K_0(r)$ and $K(r,k)$ are proportionality constants and the same relations hold for left eigenstates replacing $r\rightarrow 1/r$ into the proportionality constants. 
	
	\subsection{Edge states}
	
	Regarding the edge state we have that eq.~\eqref{propEdge} yields
	\begin{equation}
	\frac{1}{N}
	\sum_{k}
	e^{ikn}
	\sum_{s=\pm}
	\frac{f_{ab}(k)}{- 2E_s^2(k)}
	=
	K_0(r)
	\left(-\frac{c}{d}\right)^n
	r^{n-1}.
	\end{equation}
	Applying discrete Fourier transform to both sides and performing the sum in the rhs we get
	\begin{equation}
	K_0(r)=\left[c-c\left(-\frac{cr}{d}\right)^N\right]^{-1}
	\end{equation}
	which proves eqs.~\eqref{edgeSSHresult}
	
	\subsection{Bulk states}
	
	Considering the bulk states, eq.~\eqref{propBulk} yields for the A components (the same holds for B components)
	\begin{equation*}
	\frac{1}{2N}
	\sum_{k'}
	\sum_{s=\pm}
	\frac{f_{ab}(k')}{E_s(k')}
	\frac{e^{ik'n}}{\tilde E_\pm( k/2) - E_s(k')}
	=
	K(r,k)
	\alpha_{n,k}
	\end{equation*}
	Applying again discrete Fourier transform and performing the sum in the rhs we get
	\begin{equation}
	K(r,k_q/2)= \left[d(-1+(-1)^q r^N)\sin\frac{k_q}{2}\right]^{-1}
	\end{equation}
	which proves eqs.~\eqref{bulkSSHA}-\eqref{bulkSSHB}.

		\bibliography{all}
		\bibliographystyle{apsrev4-1}

\end{document}